\documentclass[aip,reprint,jcp,twocolumn]{revtex4-1}
\usepackage[paperheight=11in,paperwidth=8.25in,margin=0.67in]{geometry}
\usepackage{amsmath}
\usepackage{graphicx,color}
\usepackage{stix}

\begin{document}
\title{Lowering of the complexity of quantum chemistry methods by choice of representation}

\author{Narbe Mardirossian}
\email{nmardirossian@berkeley.edu}
\affiliation{Division of Chemistry and Chemical Engineering, California Institute of Technology, Pasadena, CA 91125}
\author{James D. McClain}
\affiliation{Division of Chemistry and Chemical Engineering, California Institute of Technology, Pasadena, CA 91125}
\author{Garnet Kin-Lic Chan}
\email{gkc1000@gmail.com}
\affiliation{Division of Chemistry and Chemical Engineering, California Institute of Technology, Pasadena, CA 91125}

\begin{abstract}
The complexity of the standard hierarchy of quantum chemistry methods is not invariant to the 
choice of representation. This work explores how the scaling of common quantum chemistry 
methods can be reduced using real-space, momentum-space, and time-dependent intermediate 
representations without introducing approximations. We find the scalings of exact Gaussian basis 
Hartree--Fock theory, second-order M{\o}ller-Plesset perturbation theory, and coupled cluster 
theory (specifically, linearized coupled cluster doubles and the distinguishable cluster approximation with doubles) 
to be $\mathcal{O}(N^3)$, $\mathcal{O}(N^3)$, and $\mathcal{O}(N^5)$ respectively, where $N$ denotes 
system size. These scalings are not asymptotic and hold over all ranges of $N$.
\end{abstract}

\maketitle

\section{Introduction}
\label{sec:introduction}

A great deal of progress in quantum chemistry comes from introducing approximations, for instance, 
to the structure of the wavefunction. For the conventional ladder of quantum chemistry methods 
(i.e., mean-field theory, perturbation theory, coupled cluster theory, etc.) such approximations 
lead to significant reductions in cost relative to the formal scaling of the methods. For example, 
within a Gaussian basis, the exact scaling of Hartree--Fock theory (HF), second-order 
M{\o}ller-Plesset perturbation theory (MP2), and coupled cluster theory with (singles and) doubles 
(CC(S)D), is commonly accepted to be $\mathcal{O}(N^4)$, $\mathcal{O}(N^5)$, and $\mathcal{O}(N^6)$, 
respectively, as a function of system size, $N$. However, by assuming locality in the wavefunction 
solutions, one can reduce the scaling of these methods to 
$\mathcal{O}(N)$~\cite{Schwegler1996,Challacombe1997,Ochsenfeld1998,Schuetz1999,Scuseria1999,Schuetz2001,Saeboe2001,Pinski2015}. 
Similarly, tensor factorization (i.e., density fitting, Cholesky decomposition, orbital-specific corrections and pair natural orbitals, tensor 
hypercontraction, etc.)~\cite{Meyer1973,Sierka2003,Werner2003,Koch2003,Aquilante2007,Neese2009,Yang2011,Hohenstein2012} and stochastic 
methods~\cite{Thom2007,Willow2012,Neuhauser2013,Ge2014} can yield reduced costs under different 
sets of assumptions and guarantees. For example, factorization methods exploit low-rank in either 
the solutions or the Hamiltonian, while stochastic methods exchange a deterministic guarantee of 
error for a probabilistic guarantee of variance.

In this short note, we will be concerned with an alternate strategy to reduce the cost of quantum 
chemistry methods. In particular, we will examine how we can change the complexity of a method simply 
by changing the underlying intermediate representations. While the choices of representations and 
approximations are commonly considered together, here we draw a distinction between the complexity 
lowering achieved through representation and that achieved through approximation. This is because 
changing representation does not itself introduce assumptions into the structure of the solutions, 
and in this sense, keeps the methods exact. To illustrate succinctly how representations yield a 
change in complexity while preserving exactness, consider the electronic Hamiltonian in three 
different bases: a general orbital basis, a plane-wave basis, and a real-space basis such as a grid,
\begin{align}
H &= \sum_{ij} t_{ij} a^\dag_i a_j + \sum_{ijkl} v_{ijkl} a^\dag_i a^\dag_j a_k a_l \label{eq:genorb} \\
H &= \sum_{k_1 k_2} t_{k_1 k_2} a^\dag_{k_1} a_{k_2} + \sum_{k_1 k_2 k_3} v_{k_1 k_2 k_3 K} a^\dag_{k_1} a^\dag_{k_2} a_{k_3} a_K \notag \\ & (K + k_3 = k_1 + k_2) \label{eq:planewave} \\
H &= \sum_{rr'} t_{rr'} a^\dag_r a_{r'} + \sum_{rr'} V_{rr'} n_r n_{r'}. \label{eq:realspace}
\end{align}
Each representation is exact in the sense that no system-specific structure in the matrix elements is assumed, 
but the number of elements is $\mathcal{O}(N^4)$, $\mathcal{O}(N^3)$, and $\mathcal{O}(N^2)$, 
respectively, without further approximations.

Using similar ideas, we will explore how a choice of representation affects the standard hierarchy 
of electronic structure methods. Assuming Coulombic interactions between particles, we find that 
the exact scaling of common Gaussian basis methods is $\mathcal{O}(N^3)$ for Hartree--Fock, 
$\mathcal{O}(N^3)$ for MP2, $\mathcal{O}(N^5)$ for linearized coupled cluster 
doubles~\cite{Bartlett1981,Taube2009} (LCCD), and $\mathcal{O}(N^5)$ for the distinguishable cluster 
approximation with doubles~\cite{Kats2013} (DCD). These scalings are not asymptotic but hold over 
any range of the system size, $N$. To reveal these scalings, we employ real-space, momentum-space, 
and time-dependent intermediate representations. None of these intermediate representations are 
new, and many elements of our arguments are well known from the approximation literature. However, 
we will cleanly draw a line between the mathematical operations that retain exactness of the 
methods, and those that introduce assumptions into the solutions. In this way, the scalings we 
derive are clearly free from approximation.

\section{Hartree--Fock Theory}
\label{sec:motivation}

As a warmup exercise to see how our results arise, consider the Hartree--Fock exchange energy. 
The conventional $\mathcal{O}(N^4)$ scaling of exact Hartree--Fock arises from the evaluation of 
all $\mathcal{O}(N^4)$ electron repulsion integrals, which are subsequently contracted into 
the one-particle density matrix, $\gamma$. However, at a more basic level, the Hartree--Fock 
exchange energy is simply a double integral,
\begin{align}
\label{eq:hfenergy}
E_{\text{HF}} = \int \int \frac{|\gamma(\mathbf{r_1},\mathbf{r_2})|^2}{|\mathbf{r_1}-\mathbf{r_2}|} \ d\mathbf{r_1} \ d\mathbf{r_2}.
\end{align}
Given the integrand, this integral can be ``exactly'' evaluated by quadrature with $\mathcal{O}(N^2)$ cost, 
regardless of the form of $\gamma$. 
To obtain the integrand, we must evaluate $\gamma$ (here expanded in a Gaussian basis) at the coordinates, ($\mathbf{r_1}, \mathbf{r_2}$). 
From $\gamma(\mathbf{r_1},\mathbf{r_2}) = \sum_i \phi_i^*(\mathbf{r_1}) \phi_i(\mathbf{r_2})$, we see that 
this carries $\mathcal{O}(N^3)$ cost; thus the full cost of evaluating the exchange energy is $\mathcal{O}(N^3)$.

We can also consider the cost of obtaining the Hartree--Fock solution. Hartree--Fock theory is a 
variational theory, and we can use the cost of evaluating the Lagrangian derivative as a proxy for 
the cost of solving the equations. Since the Lagrangian is an algebraic function of the variational 
parameters in the density matrix, the rules of adjoint differentiation~\cite{rall1981} dictate that the cost of the 
derivative is also $\mathcal{O}(N^3)$. Thus solving the Hartree--Fock equations (for a fixed 
number of derivative steps) is also $\mathcal{O}(N^3)$ cost.

The $\mathcal{O}(N^3)$ scaling of Hartree--Fock is certainly not a new result: it was already well-known 
from work on pseudo-spectral (PS) methods~\cite{Friesner1985,Ringnalda1990,Murphy1995}. The above 
exercise merely emphasizes that the $\mathcal{O}(N^3)$ scaling does not arise from any approximations, but 
only from the intermediate representation. Thus it describes the complexity of the exact method.

\section{Second-Order M{\o}ller--Plesset Perturbation Theory}
\label{sec:MP2}

Second-order M{\o}ller--Plesset perturbation theory (MP2) is perhaps the simplest route to including 
electron correlation effects. The conventional $\mathcal{O}(N^5)$ formal scaling arises from the 
atomic orbital to molecular orbital integral transformation necessary to evaluate the MP2 energy in a canonical 
basis. In Ref.~\citenum{Schaefer2017}, Kresse et al. showed that the algorithm could be exactly 
reformulated through a choice of representation to have only quartic scaling~\cite{Schaefer2017}. 
Here we show that the scaling of exact MP2 can be further reduced to $\mathcal{O}(N^3)$.

Our treatment of MP2 resembles that of HF in Section \ref{sec:motivation} in that we first express 
the MP2 energy as a multi-dimensional integral. We employ the time-dependent (i.e., Laplace 
transform~\cite{Haeser1992}) representation of the MP2 energy, which rewrites the energy as 
an integral over space and imaginary time. The Laplace transform is a common component of reduced 
scaling MP2 methods especially in conjunction with local approximations~\cite{Schuetz1999}, 
but here we emphasize that by itself it only corresponds to a choice of representation. Real-space 
representations of the MP2 energy have also been studied~\cite{Bischoff2012,Hirata2017}, and these are 
related to expressions found in the tensor hypercontraction literature via the latter's connection 
to quadrature~\cite{Hohenstein2012}.

As a single space-time integral, the two components of the MP2 energy, termed direct (MP2-J) 
and exchange (MP2-K), are
\begin{align}
E_\text{MP2-J} &= 2\int g_o(\mathbf{r_1},\mathbf{r_1'},\tau) g_o(\mathbf{r_2},\mathbf{r_2'},\tau) g_v(\mathbf{r_1'},\mathbf{r_1},\tau) \times \notag \\ &
g_v(\mathbf{r_2'},\mathbf{r_2},\tau) v(|\mathbf{r_1}-\mathbf{r_2}|) v(|\mathbf{r_1'}-\mathbf{r_2'}|) \ d\mathbf{R} \ d\tau \label{eq:MP2J} \\
E_\text{MP2-K} &= -\int g_o(\mathbf{r_1},\mathbf{r_2'},\tau) g_o(\mathbf{r_2},\mathbf{r_1'},\tau) g_v(\mathbf{r_1'},\mathbf{r_1},\tau) \times \notag \\ &
g_v(\mathbf{r_2'},\mathbf{r_2},\tau) v(|\mathbf{r_1}-\mathbf{r_2}|) v(|\mathbf{r_1'}- \mathbf{r_2'}|) \ d\mathbf{R} \ d\tau, \label{eq:MP2K}
\end{align}
where $d\textbf{R}$ denotes an integration over all spatial coordinates, $v(|\mathbf{r}-\mathbf{r'}|)$ 
is the Coulomb operator, and $g_o(\mathbf{r},\mathbf{r'},\tau)$ and $g_v(\mathbf{r},\mathbf{r'},\tau)$ 
are occupied and virtual Green's functions, respectively, defined as 
\begin{align}
g_o(\mathbf{r},\mathbf{r'},\tau) &= \sum_i \phi_i^{*}(\mathbf{r}) \phi_i(\mathbf{r'}) e^{-\epsilon_i \tau} \label{eq:gocc} \\
g_v(\mathbf{r},\mathbf{r'},\tau) &= \sum_a \phi_a^{*}(\mathbf{r}) \phi_a(\mathbf{r'}) e^{\epsilon_a \tau} \label{eq:gvirt}.
\end{align}
To obtain the appropriate scaling of the algorithm, it is necessary to treat the convolution 
integrals with the Coulomb operator in special way. Within the Fourier representation, using a uniform 
mesh in real and momentum space, the well-known result is that the Coulomb potential, 
$J(\mathbf{r'})$, corresponding to a charge distribution, $\rho(\mathbf{r})$, 
\begin{align}
J(\mathbf{r'}) = \int \frac{\rho(\mathbf{r})}{|\mathbf{r}-\mathbf{r'}|} \ d\mathbf{r} = (2 \pi)^{-3} \int \frac{4\pi}{G^2} \rho(\mathbf{G}) e^{-i \mathbf{G} \cdot \mathbf{r'}} \ d\mathbf{G}, \label{eq:Coul}
\end{align}
can be computed using the Fast Fourier transform with $\mathcal{O}(N \log N)$ cost, which we will 
consider $\mathcal{O}(N)$ for simplicity. Errors due to periodic images can either be thought of 
as arising from the limits of integration in the quadrature, or can be eliminated by truncating 
the Coulomb operator~\cite{Fuesti-Molnar2002,Fuesti-Molnar2003}. Alternatively, one can compute 
the Coulomb potential on unstructured grids by solving the real-space Poisson equation with 
$\mathcal{O}(N)$ cost~\cite{White1989}.

In either case, assuming that the Coulomb operator can be applied at $\mathcal{O}(N)$ cost, and 
assuming that the Green's functions have been formed (which requires the same $\mathcal{O}(N^3)$ 
operation for both MP2 components), we can break down the evaluation of the MP2-J expression into 
the following steps: 
\begin{align}
f(\mathbf{r},\mathbf{r'},\tau) &= g_o(\mathbf{r},\mathbf{r'},\tau) g_v(\mathbf{r'},\mathbf{r},\tau) \label{eq:MP2Jderiv1} \\
F(\mathbf{r'},\mathbf{r''},\tau) &= \int f(\mathbf{r},\mathbf{r'},\tau) v(|\mathbf{r}-\mathbf{r''}|) \ d\mathbf{r} \label{eq:MP2Jderiv2} \\
E_\text{MP2-J} &= 2 \int F(\mathbf{r'},\mathbf{r''},\tau) F(\mathbf{r''},\mathbf{r'},\tau) \ d\mathbf{R} \ d\tau \label{eq:MP2Jderiv3}.
\end{align}
In the first step, the two pairs of occupied and virtual Green's functions that depend on the same 
real-space indices are combined at $\mathcal{O}(N^2)$ cost, while the second step scales as 
$\mathcal{O}(N^2)$ because it involves application of the Coulomb operator at every point 
$\mathbf{r'}$. Finally, the energy evaluation is a double integral and is thus of $\mathcal{O}(N^2)$ 
cost. Note that the latter result indicates that certain variants of MP2, such as scaled 
opposite-spin MP2~\cite{Jung2004}, have an exact complexity of $\mathcal{O}(N^2)$ (aside from the 
formation of the Green's functions).

The complexity of the MP2-K expression can be determined in a similar way. We group the expressions as follows: 
\begin{gather}
G(\mathbf{r_1'},\mathbf{r_2'},\tau) = \int \! d\mathbf{r_2} \Bigl[ \int \! d\mathbf{r_1} \Bigl( g_o(\mathbf{r_1},\mathbf{r_2'},\tau) g_v(\mathbf{r_1'},\mathbf{r_1},\tau) \times \notag \\ 
v(|\mathbf{r_1}-\mathbf{r_2}|) \Bigl) g_o(\mathbf{r_2},\mathbf{r_1'},\tau) g_v(\mathbf{r_2'},\mathbf{r_2},\tau) \Bigl] \label{eq:MP2Kderiv1} \\
E_\text{MP2-K} = -\int G(\mathbf{r_1},\mathbf{r_2},\tau) v(|\mathbf{r_1}-\mathbf{r_2}|) \ d\mathbf{R} \ d\tau \label{eq:MP2Kderiv2}.
\end{gather}
With MP2-K, the four Green's functions have unique pairs of indices and cannot be straightforwardly 
combined as in MP2-J. The first step above is the most expensive, as the convolution integral 
($\mathcal{O}(N)$ cost) is carried out for the $\mathcal{O}(N^2)$ pairs of grid points. Thus the 
entire MP2 energy can be determined at $\mathcal{O}(N^3)$ cost.

\begin{figure}
\includegraphics[scale=0.53,bb=0 0 454 356]{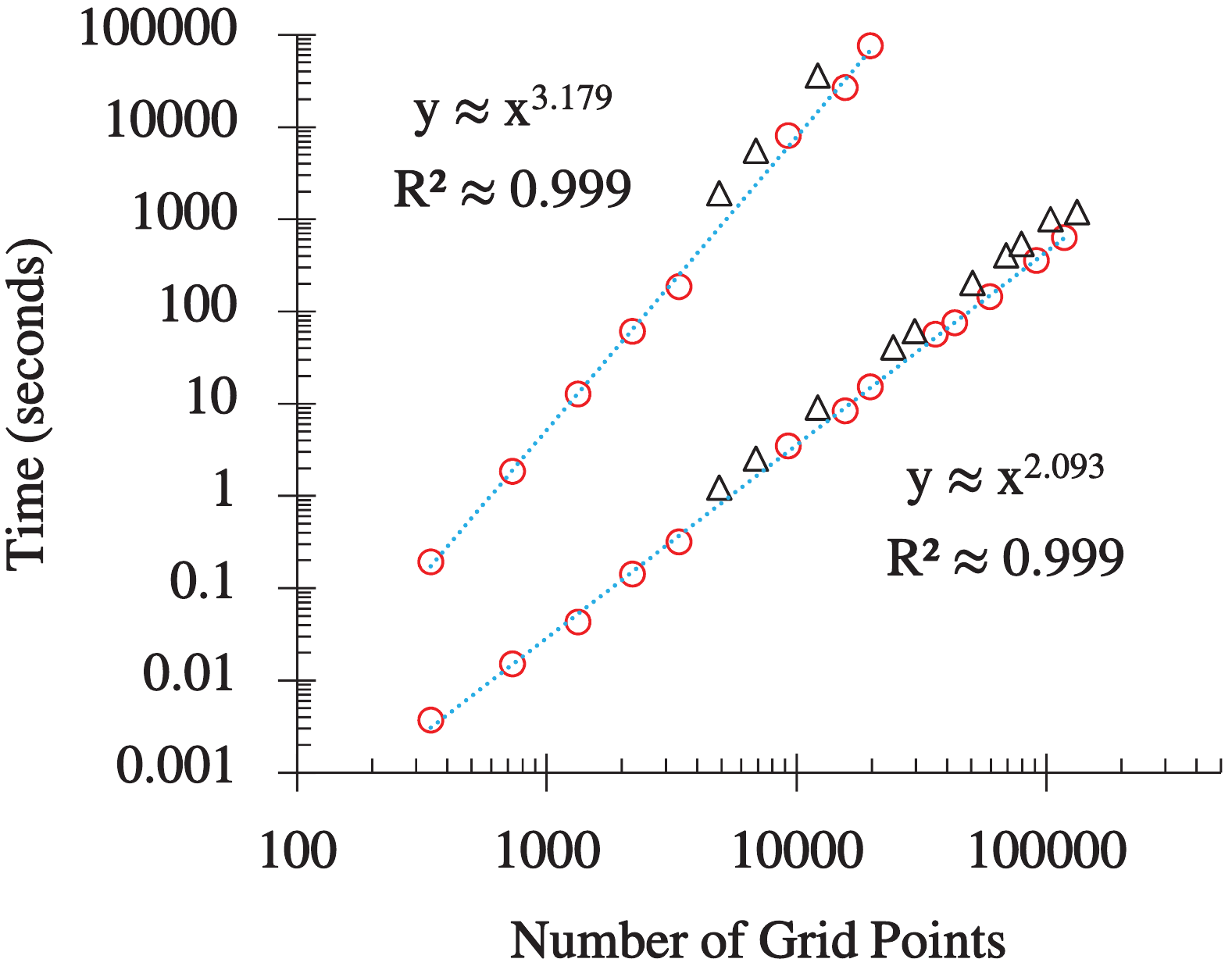}
\caption{Scaling for MP2-J (bottom) and MP2-K (top) within our pilot $\mathcal{O}(N^3 \log N)$ MP2 implementation. The circles correspond to cubic grids with a length that is a factor of 2, 3, 5, 7, 11, or 13, while the triangles correspond to prime number or non-ideal lengths.}
\label{fig:scaling}
\end{figure}

As a simple numerical demonstration of this algorithm, we have implemented an elementary cubic-scaling 
MP2 using the \textsc{PySCF} programming framework~\cite{Sun}. We start from the integral 
expressions in Eqs.~(\ref{eq:MP2J}) and (\ref{eq:MP2K}) and build the intermediates in 
Eqs.~(\ref{eq:gocc})--(\ref{eq:MP2Kderiv2}) on a uniform cubic grid. The Coulomb operator is 
applied using a three-dimensional Fast Fourier transform. Instead of the scaling with system 
size, we here carry out the simpler test of scaling with respect to the number of cubic grid 
points, which for fixed accuracy is proportional to system size. For the cubic diamond 
primitive cell (lattice constant of 6.74 Bohr, GTH-SZV basis set~\cite{VandeVondele2007} and 
GTH LDA pseudopotential~\cite{Hartwigsen1998}), the timings and scalings are shown in 
Figure \ref{fig:scaling} for a single Laplace point evaluation. We see clearly that the MP2-J 
algorithm scales close to quadratically with the number of grid points (its formal scaling 
in the implementation is $\mathcal{O}(N^2 \log N)$), while the MP2-K algorithm scales close 
to cubically with the number of grid points. The cubic scaling and conventional evaluation 
of the Laplace transform MP2 energy agree to 12 significant figures.

\section{Coupled Cluster Theory}
\label{sec:CC}

The above general arguments can be repeated to derive lower formal complexities for a variety 
of different quantum chemistry methods. Here we will briefly outline how they can be extended 
to several coupled cluster approximations. Unlike in MP2, the coupled cluster amplitudes are not 
known explicitly but must be determined by solving the amplitude equations. For simplicity, we 
will discuss only the case of CCD (the singles contribution is subleading in complexity), where 
the $t_2$ amplitude is the four index tensor $t_{ij}^{ab}$. 
Conventionally, the cost of CCD is considered to be $\mathcal{O}(N^6)$. 
Here we show that certain subsets of diagrams that have $\mathcal{O}(N^6)$ cost 
(the LCCD and DCD subsets) can be reduced to $\mathcal{O}(N^5)$ cost 
without assuming any structure in the amplitudes. A similar asymptotic scaling in a plane-wave 
basis, using a tensor hypercontraction approximation for the integrals but also without assuming 
structure in the amplitudes, has recently been reported in Ref.~\citenum{Hummel2017}. Our 
analysis is related to that in Ref.~\citenum{Hummel2017}, but illustrates that the 
$\mathcal{O}(N^5)$ scaling is an exact, rather than asymptotic, result.

The coupled cluster doubles correlation energy is given by the trace of the amplitudes with the 
integrals in Eq.~(\ref{eq:genorb}) (assuming spin orbitals),
\begin{align}
E_\text{CCD} &= \frac{1}{4} \sum_{ijab} t_{ij}^{ab} v_{ijab} \notag \\
&= \frac{1}{4} \int t({\mathbf{r_1},\mathbf{r_2},\mathbf{r_1},\mathbf{r_2}}) v(|\mathbf{r_1}-\mathbf{r_2}|) \ d\mathbf{r_1} \ d\mathbf{r_2} \label{eq:energy}, 
\end{align}
where the real space amplitude is defined as
\begin{align}
t({\mathbf{r_1},\mathbf{r_2},\mathbf{r_1'},\mathbf{r_2'}}) &= \sum_{ijab} t_{ij}^{ab} [ \phi_i^{*}(\mathbf{r_1}) \phi_j^{*}(\mathbf{r_2}) \phi_a(\mathbf{r_1'}) \phi_b(\mathbf{r_2'}) \notag \\
& - \phi_i^{*}(\mathbf{r_1}) \phi_j^{*}(\mathbf{r_2}) \phi_b(\mathbf{r_1'}) \phi_a(\mathbf{r_2'}) ].
\end{align}
The coupled cluster energy is a double integral and thus given the real-space amplitudes, 
requires $\mathcal{O}(N^2)$ cost. However, the amplitude equations do not define the amplitudes 
in this form, and the transformation from the orbital basis to real-space is of 
$\mathcal{O}(N^5)$ cost. Thus the exact cost to evaluate the coupled cluster energy is 
$\mathcal{O}(N^5)$, without further assumptions.

\begin{figure*}
\includegraphics[scale=0.54,bb=0 0 784 546]{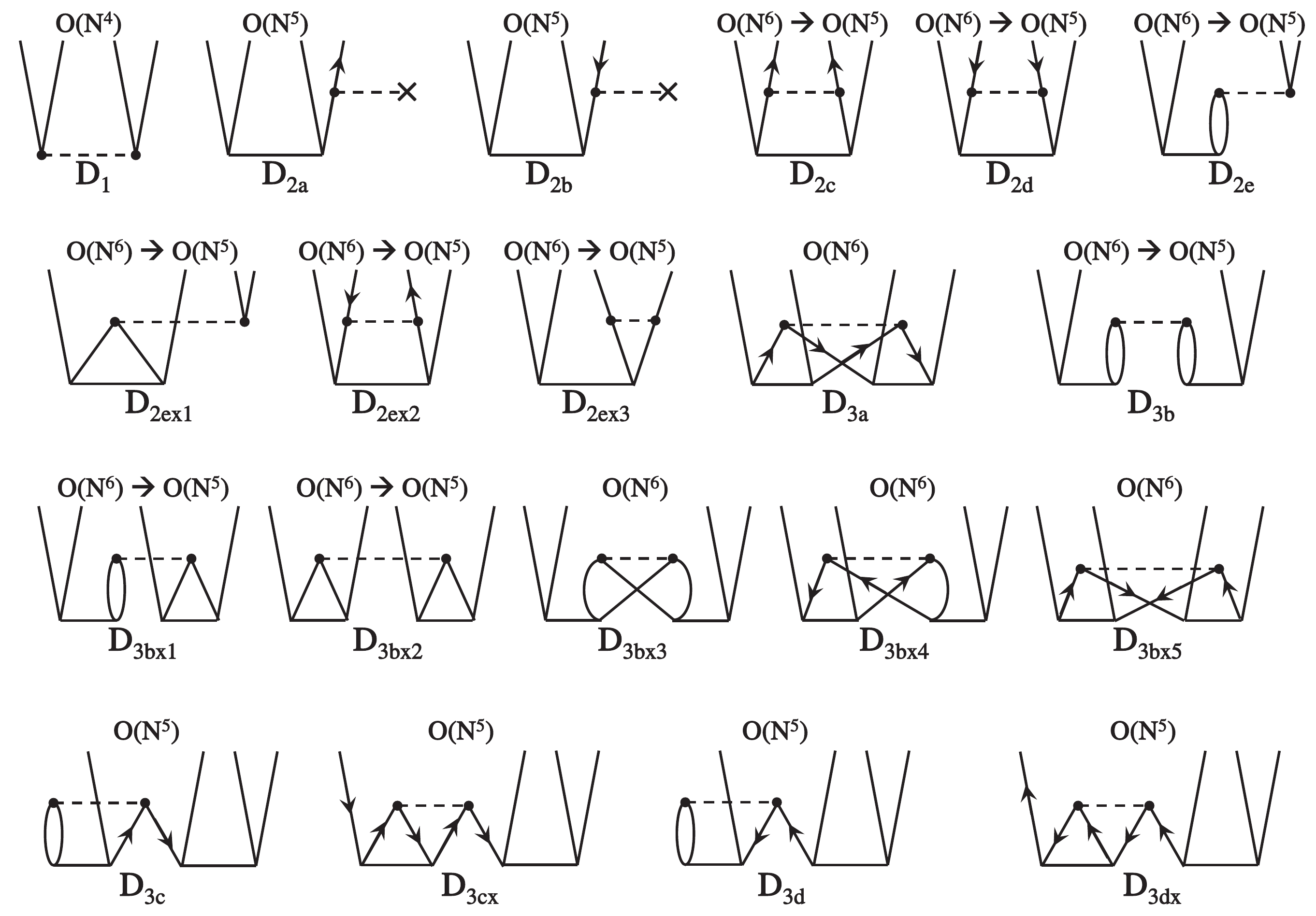}
\caption{Spin-summed CCD amplitude equation diagrams. The cost of diagrams associated with a 
single scaling is unchanged when using a real-space intermediate representation, 
while those with arrows experience scaling reduction (i.e., the scaling of term D$_{2c}$ is 
reduced from $\mathcal{O}(N^6)$ to $\mathcal{O}(N^5)$).}
\label{fig:ccd}
\end{figure*}

The CCD amplitude equations~\cite{shavitt2009many} are conveniently presented in diagrammatic form in Figure \ref{fig:ccd}. 
Above each diagram, we give the scaling of each term. Like the above argument for the 
energy, we can transform the indices of each amplitude into the real-space representation as 
needed to apply the Coulomb operator, before transforming back into the orbital basis. This 
change of representation reduces the complexity of nine of the 20 terms from $\mathcal{O}(N^6)$ to 
$\mathcal{O}(N^5)$. These reductions are also indicated in Figure \ref{fig:ccd}. According to the 
diagrams, LCCD corresponds to the first nine terms, while DCD corresponds to LCCD, plus the three 
other terms whose scaling is reduced from $\mathcal{O}(N^6)$ to $\mathcal{O}(N^5)$ (D$_{3b}$, D$_{3bx1}$, and D$_{3bx2}$), plus the last 
four $\mathcal{O}(N^5)$ terms. Thus the exact cost to determine the amplitudes in either LCCD or 
DCD is $\mathcal{O}(N^5)$.
\begin{gather}
\frac{1}{2} \sum_{CD} \langle AB | \hat{v} | CD \rangle t_{IJ}^{CD} \notag \\
\frac{1}{2} \sum_{CD} \int A(\mathbf{r_1}) B(\mathbf{r_2}) v(\mathbf{r_1},\mathbf{r_2}) C(\mathbf{r_1}) D(\mathbf{r_2}) t_{IJ}^{CD} \ d\mathbf{r_1} \ d\mathbf{r_2} \notag \\
t_{IJ}^{\phantom{C}D}(\mathbf{r_1})=\sum_C C(\mathbf{r_1}) t_{IJ}^{CD} \ \ \ \ \ \ \ t_{IJ}(\mathbf{r_1},\mathbf{r_2})=\sum_D D(\mathbf{r_2}) t_{IJ}^{\phantom{C}D}(\mathbf{r_1}) \notag \\
T_{IJ}(\mathbf{r_1},\mathbf{r_2})=t_{IJ}(\mathbf{r_1},\mathbf{r_2}) v(\mathbf{r_1},\mathbf{r_2}) \label{eq:ccdproof1} \\
T_{IJ}^A(\mathbf{r_2})=\int A(\mathbf{r_1}) T_{IJ}(\mathbf{r_1},\mathbf{r_2}) \ d\mathbf{r_1} \notag \\
T_{IJ}^{AB}=\frac{1}{2} \int B(\mathbf{r_2}) T_{IJ}^A(\mathbf{r_2}) \ d\mathbf{r_2} \notag
\end{gather}
The nine diagrams with reduced complexity can be grouped into three separate types: 1) D$_{2c}$, 
D$_{2d}$, D$_{2ex2}$, and D$_{2ex3}$, 2) D$_{2e}$ and D$_{2ex1}$, and 3) D$_{3b}$, D$_{3bx1}$, and 
D$_{3bx2}$. The type 1 terms contain a single $t_2$ amplitude, with the contraction indices 
corresponding to different electron coordinates (i.e., $\mathbf{r_1}$ and $\mathbf{r_2}$), 
while the type 2 terms contain a single $t_2$ amplitude, with the contraction indices corresponding 
to the same electron coordinate (i.e., either both $\mathbf{r_1}$ or both $\mathbf{r_2}$). The 
type 3 terms, despite containing a pair of $t_2$ amplitudes, can be evaluated in $\mathcal{O}(N^5)$ 
time because the contraction indices contained in each amplitude correspond to the same 
electron coordinate. To illustrate the scaling reduction for the three aforementioned types, 
we take a single term from each case and define appropriate intermediates in Eqs.~\ref{eq:ccdproof1}--\ref{eq:ccdproof3}, 
where Eq.~\ref{eq:ccdproof1} corresponds to diagram D$_{2c}$, Eq.~\ref{eq:ccdproof2} corresponds to diagram D$_{2e}$, 
and Eq.~\ref{eq:ccdproof3} corresponds to diagram D$_{3b}$.
\begin{gather}
2 \sum_{KC} \langle KB | \hat{v} | CJ \rangle t_{IK}^{AC} \notag \\
2 \sum_{KC} \int K(\mathbf{r_1}) B(\mathbf{r_2}) v(\mathbf{r_1},\mathbf{r_2}) C(\mathbf{r_1}) J(\mathbf{r_2}) t_{IK}^{AC} \ d\mathbf{r_1} \ d\mathbf{r_2} \notag \\
t_{I\phantom{K}}^{AC}(\mathbf{r_1})=\sum_K K(\mathbf{r_1}) t_{IK}^{AC} \ \ \ \ \ \ \ t_{I}^{A}(\mathbf{r_1})=\sum_C C(\mathbf{r_1}) t_{I\phantom{K}}^{AC}(\mathbf{r_1}) \notag \\
T_{I}^{A}(\mathbf{r_2})=\int t_{I}^{A}(\mathbf{r_1}) v(\mathbf{r_1},\mathbf{r_2}) \ d\mathbf{r_1} \label{eq:ccdproof2} \\
T_{IJ}^{AB}=2 \int B(\mathbf{r_2}) J(\mathbf{r_2}) T_{I}^{A}(\mathbf{r_2}) \ d\mathbf{r_2} \notag
\end{gather}
In order to clarify the reduction in scaling, we will walk through the derivation for the D$_{2c}$ term. 
In a Gaussian basis, it is evident that this term scales as $\mathcal{O}(N^6)$. After rewriting the integral in 
its real-space form, the contractions over C and D each require $\mathcal{O}(N^5)$ time, since the former 
involves four orbital indices and one real-space index, and the latter involves three orbital indices and two real-space indices. 
Then, the result is multiplied by the Coulomb operator in real space, at $\mathcal{O}(N^4)$ cost. The next step is similar 
to that shown in Equation \ref{eq:Coul}, and scales as $\mathcal{O}(N^3)$, while the final step is again 
$\mathcal{O}(N^5)$. Thus, the scaling for a term that is conventionally $\mathcal{O}(N^6)$ can be exactly 
reduced to $\mathcal{O}(N^5)$.
\begin{gather}
2 \sum_{KLCD} \langle KL | \hat{v} | CD \rangle t_{IK}^{AC} t_{LJ}^{DB} \notag \\
2 \sum_{KLCD} \int K(\mathbf{r_1}) L(\mathbf{r_2}) v(\mathbf{r_1},\mathbf{r_2}) C(\mathbf{r_1}) D(\mathbf{r_2}) t_{IK}^{AC} t_{LJ}^{DB} \ d\mathbf{r_1} \ d\mathbf{r_2} \notag \\
t_{I\phantom{K}}^{AC}(\mathbf{r_1})=\sum_K K(\mathbf{r_1}) t_{IK}^{AC} \ \ \ \ \ \ \ t_{I}^{A}(\mathbf{r_1})=\sum_C C(\mathbf{r_1}) t_{I\phantom{K}}^{AC}(\mathbf{r_1}) \notag \\
t_{\phantom{L}J}^{DB}(\mathbf{r_2})=\sum_L L(\mathbf{r_2}) t_{LJ}^{DB} \ \ \ \ \ \ \ t_{J}^{B}(\mathbf{r_2})=\sum_D D(\mathbf{r_2}) t_{\phantom{L}J}^{DB}(\mathbf{r_2}) \notag \\
T_{I}^{A}(\mathbf{r_2})=\int t_{I}^{A}(\mathbf{r_1}) v(\mathbf{r_1},\mathbf{r_2}) \ d\mathbf{r_1} \label{eq:ccdproof3} \\
T_{IJ}^{AB}=2 \int t_{J}^{B}(\mathbf{r_2}) T_{I}^{A}(\mathbf{r_2}) \ d\mathbf{r_2} \notag
\end{gather}

\section{Alternate Representations}
\label{sec:AOMP2}

In our above arguments, we reduced the exact scalings of quantum chemistry methods by combining 
several different representations. In all three methods (HF, MP2, CC), we used a real-space 
intermediate representation. We obtained additional cost reductions from the Fourier representation 
of the Coulomb operator, while the MP2 algorithm also used a time-dependent representation. These 
intermediate representations are not the only ones that lead to reduced scalings, and other 
choices may lead to lower computational prefactors. For example, if we allow for a polylogarithmic 
dependence of computational cost on a threshold error $\epsilon$, then we can regard atomic orbital 
bases as a form of exponentially localized real-space basis. This is the standard argument for atomic 
orbital screening, but here we are interested only in the reduction in complexity that can be 
achieved without assuming locality in the wavefunction or by cutting off algebraically decaying 
quantities~\cite{Wilson1997,Doser2009}. Within this sense of retaining the exactness of the method, 
as long as we also use a scheme to apply the Coulomb operator with $\mathcal{O}(N)$ cost, one 
recovers the same complexities we have derived above, for systems on length scales larger than the 
atomic orbital size.

One way to apply the Coulomb operator in a fast scheme is to use a mixed basis and grid representation, 
as is commonly done in mixed Gaussian and plane-wave implementations~\cite{Fuesti-Molnar2002,Hutter2014,McClain2017,Sun2017a} 
where the Coulomb operator is applied, as above, in the Fourier representation. As an 
explicit example, we outline how to evaluate the MP2-J term using this idea as well as 
an atomic orbital representation. Here we use the standard Roman and Greek symbols for 
molecular orbitals and atomic orbitals respectively, with the molecular orbitals expanded 
as $\phi_p(\mathbf{r}) = \sum_{\mu} C_{\mu p} \mu(\mathbf{r})$. Contributions of atomic 
orbital products $\mu(\mathbf{r})\nu(\mathbf{r})$ will be assumed screened if $||\mu \nu|| < \epsilon$, 
and screened pairs will be indicated by the symbol $\langle \mu\nu\rangle$. 
Starting with the atomic orbital Laplace transform expression for MP2-J, 
\begin{align}
E_\text{MP2-J} = 2\int \sum_{\mu' \nu' \sigma' \lambda'} \sum_{\mu \nu \sigma \lambda} (\mu' \nu' | \sigma' \lambda') (\nu \mu | \lambda \sigma) \times \notag \\
g_{\mu' \mu} g_{\sigma' \sigma} \bar{g}_{\nu' \nu} \bar{g}_{\lambda' \lambda} \ d\tau \label{eq:AOMP2J},
\end{align}
with the atomic orbital Green's functions (that require $\mathcal{O}(N^3)$ time to compute) defined as
\begin{align}
g_{\mu' \mu}(\tau) &= \sum_i C_{\mu' i} C_{\mu i} e^{-\epsilon_i \tau} \\
\bar{g}_{\nu' \nu}(\tau) &= \sum_a C_{\nu' a} C_{\nu a} e^{\epsilon_a \tau},
\end{align}
it is possible to formulate a series of steps to evaluate Eq.~(\ref{eq:AOMP2J}) where the cost is 
no greater than $\mathcal{O}(N^3)$. The first three intermediates require $\mathcal{O}(N^3)$ cost, 
\begin{align}
\rho_{\mu}(\mathbf{r},\tau) &= \sum_{\mu'} \mu'(\mathbf{r}) g_{\mu' \mu}(\tau) \\
h_{\langle \mu \nu \rangle}(\mathbf{r},\tau) &= \int \rho_{\mu}(\mathbf{r'}) \rho_{\nu}(\mathbf{r'}) v(|\mathbf{r}-\mathbf{r'}|) \ d\mathbf{r'} \\
h_{\langle \mu \nu \rangle \langle \sigma' \lambda' \rangle}(\tau) &= \int h_{\langle \mu \nu \rangle}(\mathbf{r},\tau) \langle \sigma' \lambda' \rangle(\mathbf{r}) \ d\mathbf{r}, \\
\end{align}
and the final energy evaluation, 
\begin{align}
E_\text{MP2-J} = 2\int \sum_{\langle \sigma' \lambda' \rangle,\langle \mu \nu \rangle} h_{\langle \mu \nu \rangle \langle \sigma' \lambda' \rangle}(\tau) h_{\langle \lambda' \sigma' \rangle \langle \nu \mu \rangle}(\tau) \ d\tau,
\end{align}
is $\mathcal{O}(N^2)$ cost. 
Note that there are additional cubic steps in the evaluation of the MP2-J term compared to 
the quadrature implementation, because in the case of quadrature the cubic cost is confined 
to the formation of the Green's functions at the beginning of the algorithm, while here the cubic 
cost is delayed until quantities are placed on the grid. However, it is clear that by using 
a mixture of atomic orbitals and quadrature, the overall prefactor is greatly reduced, as typically the number 
of atomic orbitals required in the ``function'' quadrature is much less than the number of 
grid points required for numerical quadrature. Although the above algorithm will exhibit cubic 
scaling on length scales determined only by the atomic orbitals rather than the locality of 
the wavefunction, the use of diffuse functions will prevent the onset of this scaling until 
larger systems. In such a case, alternative representations may prove useful, and this is a 
topic of future work. 

\section{Conclusion}
\label{sec:conc}

In summary, the present work re-examines the exact scaling of several traditional quantum 
chemistry methods. We find that the freedom of choice of intermediates means that HF and 
MP2 can be reduced to cubic scaling, and variants of coupled cluster such as linearized 
coupled cluster doubles (LCCD) and the distinguishable cluster 
approximation with doubles (DCD) may be reduced to $\mathcal{O}(N^5)$ scaling. 
These are scalings of the exact methods in the sense that no assumptions are made about the forms 
of the solutions. The exact scalings that we describe encourage a modified perspective 
on several topics. For example, they lead to a different organization of the correlation 
hierarchy, where the complexity gap between density functional methods~\cite{Mardirossian2017a} 
and traditional wavefunction methods is eliminated. They also suggest a new way to 
classify diagrams in coupled cluster theory~\cite{Bartlett2007} that may lead to new 
correlation approximations. Finally, given that the exact scalings are lower than that 
of many current approximate methods, substantial further reductions in cost can be 
obtained in practice by combining the ideas here with the rich existing set of techniques 
used to define approximate quantum chemistry methods.

\acknowledgments{This work was supported by the US National Science Foundation through NSF:CHE 1665333 and NSF:SSI 1657286. GKC is a Simons Investigator in Theoretical Physics.}

\bibliography{paper_ref_full}

\end{document}